\documentclass[epj]{svjour}
\usepackage{graphics}
\usepackage{hyperref}
\usepackage{xcolor}
\begin{document}
\title{Molecular dynamics study of accelerated ion-induced \\ shock waves in biological media}

\author{Pablo de Vera\inst{1,2,3,}\thanks{Corresponding author: p.devera@qub.ac.uk}, Nigel J. Mason\inst{2}, Fred J. Currell\inst{1}, and Andrey V. Solov'yov\inst{3,}\thanks{On leave from A. F. Ioffe Physical Technical Institute, 194021 St. Petersburg, Russian Federation}
%
}                     
\institute{School of Mathematics and Physics, Queen's University Belfast, BT7 1NN Belfast, Northern Ireland, United Kingdom \and Department of Physical Sciences, The Open University, Walton Hall, MK7 6AA Milton Keynes, England, United Kingdom \and MBN Research Center, Altenh\"{o}ferallee 3, 60438 Frankfurt am Main, Germany}
\date{Received: date / Revised version: date}
%
\abstract{
We present a molecular dynamics study of the effects of carbon- and iron-ion induced shock waves in DNA duplexes in liquid water. 
We use the CHARMM force field implemented within the MBN Explorer simulation package to optimize and equilibrate DNA duplexes in liquid water boxes of different sizes and shapes. The translational and vibrational degrees of freedom of water molecules are excited according to the energy deposited by the ions and the subsequent shock waves in liquid water are simulated. The pressure waves generated are studied and compared with an analytical hydrodynamics model which serves as a benchmark for evaluating the suitability of the simulation boxes. The energy deposition in the DNA backbone bonds is also monitored as an estimation of biological damage, something which lies beyond the possibilities of the analytical model.
} 

\authorrunning{P. de Vera \textit{et al.}}
\titlerunning{Molecular dynamics study of ion-induced shock waves in biological media}

\maketitle
\section{Introduction}
\label{intro}

Collisional phenomena between fast ions and  biomolecules is a topic of major interest since it is necessary to understand the mechanisms of such processes if we are to develop ion beam  cancer therapy (IBCT). 
In this therapy technique, energetic protons or heavier ions are used clinically to treat deeply seated tumors \cite{Loeffler2013}. From a macroscopic point of view one of the working principles of IBCT is the Bragg peak, a sharp maximum in the depth-dose curve at the end of the energetic ion trajectories (contrary to photon or electron beams which have a quite broad energy deposition profile) that maximizes energy deposition in the tumor while sparing surrounding healthy tissues. However, it is well known that the effectiveness of IBCT relies on nanoscopic phenomena rather than on macroscopic characteristics \cite{Scholz2007,Surdutovich2014}, the former being directly related to atomic collisions with biomolecules. Indeed, a given dose deposited by ions presents a much larger cell killing probability than the same dose deposited by photons. This increased relative biological effectiveness is due to the large energy deposited around ion tracks on the nanoscale, giving place to an increase in the clustering of damaging events in biomolecules, especially in nuclear DNA, which makes the repair processes less effective \cite{Schardt2010}.

The fundamental aspects of the problem are also of great interest. The irradiation with ions involves new physical phenomena which are not always considered properly (or considered at all) in biophysical models. In fact, IBCT is a complex problem, involving many different space, energy, and time scales, ranging from the transport of energetic ions in macroscopic tissues, the production of secondary electrons and radicals that can propagate on the nano- and microscale (molecular and cellular levels, respectively) and their interaction with biomolecules on the nanometer scale leading to the final biological outcomes, noticeable in larger space and time scales \cite{Surdutovich2014}. More significantly, all the physico-chemical processes occurring at the molecular level make it necessary to deviate from a simple energy deposition scheme. On these space scales not only energy deposition events are of importance, since the way in which this energy promotes different kind of processes can significantly affect the final effects. While many processes are fairly well known, such as the electron production and propagation or the generation of free radicals, new interactions are being discovered, such as the dissociative electron attachment, a mechanism by which very low energy electrons (with energies even below the ionization threshold) can fragment biomolecules \cite{Boudaiffa2000}.



\begin{sloppypar}
In this context another damage mechanism has been theoretically predicted: ion-induced shock waves on the nanometer scale \cite{Surdutovich2013,Surdutovich2010,Yakubovich2012}.
Ion beams can deposit large amounts of energy per unit path length (a carbon ion in the Bragg peak region deposits 900 eV/nm), and the major part of this energy is used to eject secondary electrons of very low energies, below 50 eV \cite{Surdutovich2014,deVera2013,deVera2013b}. Most of such electrons transfer their energy to electronic excitations of the medium in less than a nanometer and the time scale in which this energy loss occurs is very short, of a few femtoseconds \cite{Surdutovich2015}. These times are very short in comparison with the mechanism capable of dissipating this energy, the electron-phonon coupling, which occurs in the sub-picosecond scale \cite{Surdutovich2015}. This situation results in a large heating of the medium in nanocylinders around the ion tracks, providing the conditions for a violent explosion of these ``hot cylinders'', a mechanism we refer to as ion-induced shock waves.
\end{sloppypar}

This shock wave effect was first predicted in terms of a hydrodynamics model where it was shown that pressures up to tens of GPa can be produced around ion tracks \cite{Surdutovich2010}. However, this is not enough for predicting biomolecular damage and subsequently molecular dynamics simulations were used to show how these conditions are sufficient to produce bond breaking in nucleosomes \cite{Surdutovich2013}. Moreover since the shock waves travel at high velocity  they can propagate secondary species (i.e., free radicals, solvated electrons) much faster than the diffusion mechanism. All these dynamical and thermo-mechanical effects can drastically change the physico-chemical environment to which biomolecules are exposed under irradiation. 



A proper understanding of the characteristics of shock waves requires more systematic studies, especially by the use of molecular dynamics, a technique that can assess their properties and consequences in more detail. Such studies will enhance the understanding of the biological role of ion-induced shock waves, as well as help in the design of possible experiments to verify their existence. In the present paper we report molecular dynamics simulations to study the main features of ion-induced shock waves and their effects on DNA by means of the MBN Explorer software \cite{Solovyov2012}.
We focus the work on shock waves produced around carbon (one of the most promising projectiles used in IBCT) and iron ion tracks in the Bragg peak region. We study the dependence of several quantities, namely pressure waves and energy deposition in DNA bonds, on the size and characteristics of the system, in order to establish the proper features of the simulation box for future more systematic studies.  These results are compared with the analytical hydrodynamics model \cite{Surdutovich2010} as well as to previous simulation results \cite{Surdutovich2013} to benchmark the simulations.
As a probe for biodamage we will use short DNA segments. This is because such short DNA duplexes can be simulated more straightforwardly than nucleosomes, so they are more convenient for systematic studies. Also, since the effects of shock waves are more noticeable in short distances, a short DNA strand should be enough for evaluating biomolecular damage, in a similar fashion to the effects of secondary electrons on DNA studied previously \cite{Surdutovich2014}.

The methodology of the work is explained in section \ref{sec:methods}, where the hydrodynamics model (subsection \ref{sec:hydrodynamics}) and the molecular dynamics procedure (subsection \ref{sec:MM}) are discussed. 
The results of the simulations are presented in section \ref{sec:results}, where the pressure generated by the shock waves and their effects in the DNA duplex are studied. The final conclusions and remarks are given in section \ref{sec:conclusions}.




\section{Methods}
\label{sec:methods}

Ion-induced nanoscopic shock waves were first described in terms of an analytical hydrodynamics model that was used for the first evaluation of their characteristics \cite{Surdutovich2010}. Even though it allows calculation of the basic physical features, such as the pressure generated during the shock wave, it does not allow the evaluation of the characteristics related to biological effects, such as the damage of DNA molecules, for which an atomic level description is needed. This kind of analysis can be performed through molecular dynamics simulations \cite{Surdutovich2013}. However, the analytical hydrodynamics model serves as a good benchmark for the molecular dynamics simulations. Some relevant features of the analytical hydrodynamics model are reviewed in subsection \ref{sec:hydrodynamics}. The molecular dynamics simulations technique is reviewed in subsection \ref{sec:MM}. The choice of the biomolecular probe, as well as its setting up for the simulations, is explained in subsection \ref{sec:system}, and the setting up of the initial conditions for the simulation of the ion-induced shock wave is described in subsection \ref{sec:initial}.

\subsection{Hydrodynamics model}
\label{sec:hydrodynamics}

This model was adapted from a classical hydrodynamics treatment of the self-similar flow of liquid water and heat transfer \cite{Landau1987} and applied to the specific situation of the energy delivered around an energetic ion track on the nanoscale where the resulting water flow is cylindrical \cite{Surdutovich2010}.

Several predictions of this model are very convenient for benchmarking the simulations performed in the present work. In particular, the position of the wave front as a function of time is given by:
\begin{equation}
R(t) = \beta \sqrt{t} \left[\frac{\left|{\rm d}T/{\rm d}s\right|}{\rho}\right]^{1/4} \mbox{ , }
\label{eq:R_t}
\end{equation}
and the pressure of the front as a function of the front radius is:
\begin{equation}
P_{\rm front}(R) = \frac{1}{2\left(\gamma+1\right)}\frac{\beta^4 \left|{\rm d}T/{\rm d}s\right|}{R^2} {\mbox .}
\label{eq:P_R}
\end{equation}
In both equations, $\rho = 1$ g/cm$^3$ is the density of unperturbed liquid water, $\gamma=C_{\rm p}/C_{\rm v} = 1.222$ for liquid water, $\left|{\rm d}T/{\rm d}s\right|$ is the stopping power (i.e., mean energy loss ${\rm d}T$ per unit path length ${\rm d}s$) of an ion in liquid water, and $\beta$ is a parameter which value for liquid water is $\beta = 0.86$ \cite{Surdutovich2010}.

\subsection{Molecular dynamics simulation}
\label{sec:MM}

In the molecular dynamics technique \cite{Allen1989} all the atoms of the system are considered and their classical trajectories are followed by computing the interaction forces between all the atoms in the system. The evolution with time of the coordinates $\vec{r_i}(t)$ of each atom $i$ of mass $m_i$ is computed for discrete time steps ${\rm d} t$, according to the Langevin equation:
\begin{equation}
m_i \frac{{\rm d}^2\vec{r}_i}{{\rm d}t^2} = \sum_{j\neq i} \vec{F}_{ij}-\frac{1}{\tau_{\rm d}}m_i \vec{v}_i+\vec{f_i} \mbox{ , }
\label{eq:MD}
\end{equation}
where $\sum_{j\neq i} \vec{F}_{ij}$ is the total force acting on atom $i$ as a consequence of its interaction with all other atoms $j$ in the system (i.e.,  Newton's second law). The second and third terms in the right hand side of Eq. (\ref{eq:MD}) correspond to the thermostat, used to keep the temperature of the system nearly constant to $T$, when coupled to a thermal bath. In the present work the Langevin thermostat is used, which exerts a viscous force on each particle of velocity $\vec{v_i}$, as well as a random force $\vec{f_i}$ which guarantees thermal equilibrium. $\tau_{\rm d}$ is the damping time of the thermostat, while $\vec{f_i}$ is a Gaussian random force with zero mean and variance $\sigma_i^2=2m_i k_{\rm B}T/\tau_{\rm d}$ with $k_{\rm B}$ being the Boltzmann's constant.

For biomolecular systems, where the structure of the molecule is determined not only by interatomic distances but also by the geometric configuration of groups of atoms due to the molecular orbital hybridization, it is common to use special forcefields describing such interactions. In the CHARMM forcefield \cite{MacKerel1998}, one of the most common ones for describing biomolecules, the force acting on the atom $i$ is obtained from the potential energy $U(\vec{R})$ as $\sum_{j\neq i} \vec{F}_{ij} = {\rm d}U(\vec{R})/{\rm d}\vec{r_i}$ which corresponds to a given set of atomic coordinates $\vec{R}$ and is expressed as a combination of energies arising from the distances between pairs of bonded atoms, the angles formed between groups of three sequentially bonded atoms, the dihedral torsion angle formed by groups of sequentially four bonded atoms, the improper angles formed between groups of atoms that should form a plane, and the nonbonded interactions represented by the pure Coulomb force and the van der Waals interaction between pairs of atoms:
\begin{eqnarray}
U(\vec{R}) & = & \sum_{\rm bonds} K_{\rm b}(b-b_0)^2 + \sum_{\rm angles} K_{\theta}(\theta-\theta_0)^2 + \\ 
& + & \sum_{\rm dihedr.} K_{\chi}\left(1+\cos{n\chi}-\delta\right) + \sum_{\rm improp.} K_{\varphi}(\varphi-\varphi_0)^2 + \nonumber \\
& + & \sum_{i}\sum_{j\neq i} \frac{q_i q_j}{\varepsilon \, r_{ij}} + \left[ \epsilon_{ij} \left(\frac{R_{{\rm min},ij}}{r_{ij}}\right)^{12} - \left(\frac{R_{{\rm min},ij}}{r_{ij}}\right)^{6} \right] \mbox{ . } \nonumber
\label{eq:MM}
\end{eqnarray}
In this equation $b$ is the bond distance between two bonded atoms, $\theta$ is the bond angle between every triplet of sequentially bonded atoms, $\chi$ is the dihedral torsion angle formed by every four atoms connected via covalent bonds and $\varphi$ is the improper torsion angle, used to maintain planarity between groups of sequentially bonded atoms; $b_0$, $\theta_0$ and $\varphi_0$ correspond to the equilibrium quantities, while $n$ and $\delta$ determine the periodicity of the dihedral interaction. $K_b$, $K_{\theta}$, $K_{\chi}$ and $K_{\varphi}$ are the corresponding force constants. The Coulomb interaction is characterized by the atomic partial charges $q_i$, the interatomic distances $r_{ij}$ and the effective dielectric constant $\varepsilon$. The van der Waals interaction is defined by a 6--12 Lennard-Jones potential with well depth $\epsilon$ and the minimum energy distance $R_{\rm min}$. All of these parameters can be obtained for many biological molecules, including nucleic acids and proteins, from the CHARMM potential \cite{MacKerel1998}. All simulations in this work have been performed using the CHARMM implementation within the MBN Explorer simulation package \cite{Solovyov2012}.

\subsubsection{Setting up the biological system}
\label{sec:system}

\begin{sloppypar}

The main constituent of living tissues is liquid water. Therefore when we refer to the ion-induced shock waves, they occur mainly in water. However, if we want to estimate possible biological effects, we should consider as a target of the shock wave some biological molecules, for example DNA. Thus in order to set up a simulation box for the present study we need to define (i) the water box in which the shock wave is going to propagate and (ii) the biological molecule that we are going to use to assess biodamage.

Point (i), even though it seems straightforward, has to be considered carefully. This is because the shock wave is a violent dynamical process and the pressure waves generated can propagate fast, reaching, under certain circumstances, the boundaries of the simulation box. Such ``boundary effects'' can produce artifacts in the simulations. These artifacts could be simply avoided by setting a extremely large water box in order to guarantee that the shock wave will not arrive to the boundaries. However, the computational cost of molecular dynamics simulations increases with the number of atoms in the simulation box, so the size of the system has to be selected properly in order to perform the simulations in reasonable times. For this reason in the present study we have constructed four different simulation boxes. 

\begin{table}
\caption{Summary of the different simulation boxes used for the shock wave simulations. The four systems have a length of 4.6 nm along the track direction $y$, where PBC are applied.}
\label{tab:systems}       
\begin{tabular}{llll}
\hline\noalign{\smallskip}
System & Shape & Track-to-boundary & PBC  \\
	   &       & distance (nm)     & ($x$, $z$) \\
\noalign{\smallskip}\hline\noalign{\smallskip}
I & disc & 10 & No \\
II & parallelepiped & 10 & Yes \\
III & disc & 17 & No \\
IV & parallelepiped & 17 & Yes \\
\noalign{\smallskip}\hline
\end{tabular}
\end{table}

In particular we want to study the effect of the size of the system in the simulation results as well as the inclusion of periodic boundary conditions to simulate extended media. These four systems are summarized in Table \ref{tab:systems}. The first box, which we will refer as system I, was designed as a liquid water disc of 10 nm radius around the track, located in the disc center. This system is quite convenient since it is quite small and then computationally efficient. A radius of 10 nm was chosen following results from the hydrodynamics model, which shows that the pressure of the shock wave is already very small after 10 nm \cite{Surdutovich2010}. This disc, 4.6 nm long in the track direction $y$, was simulated with periodic boundary conditions (PBC) only in this direction. In order to check the effect of the inclusion of PBC in the directions $x$ and $z$, the parallelepiped system II was built. It has the same length of 4.6 nm in the track direction $y$ and the distance from the track to the boundary is also 10 nm but PBC are applied in the three coordinates. Finally, to check size effects, two similar systems but with a track-to-boundary distance of 17 nm where built, systems III and IV. System IV will be in general used as a benchmark for the rest of them since it is the largest one and it has PBC in all directions.

Regarding point (ii), in the present work we would like to chose a simple system that can be simulated straightforwardly for performing more systematic studies of shock wave effects in biomolecules.
As a first approximation we will use short DNA segments, as it was done in previous works evaluating the damaging effect of secondary electrons \cite{Surdutovich2014}. They are simpler to treat than nucleosomes and, since shock wave effects are quite local, they should be large enough for estimating DNA damage.

\begin{figure}
\resizebox{0.9\columnwidth}{!}{%
  \includegraphics{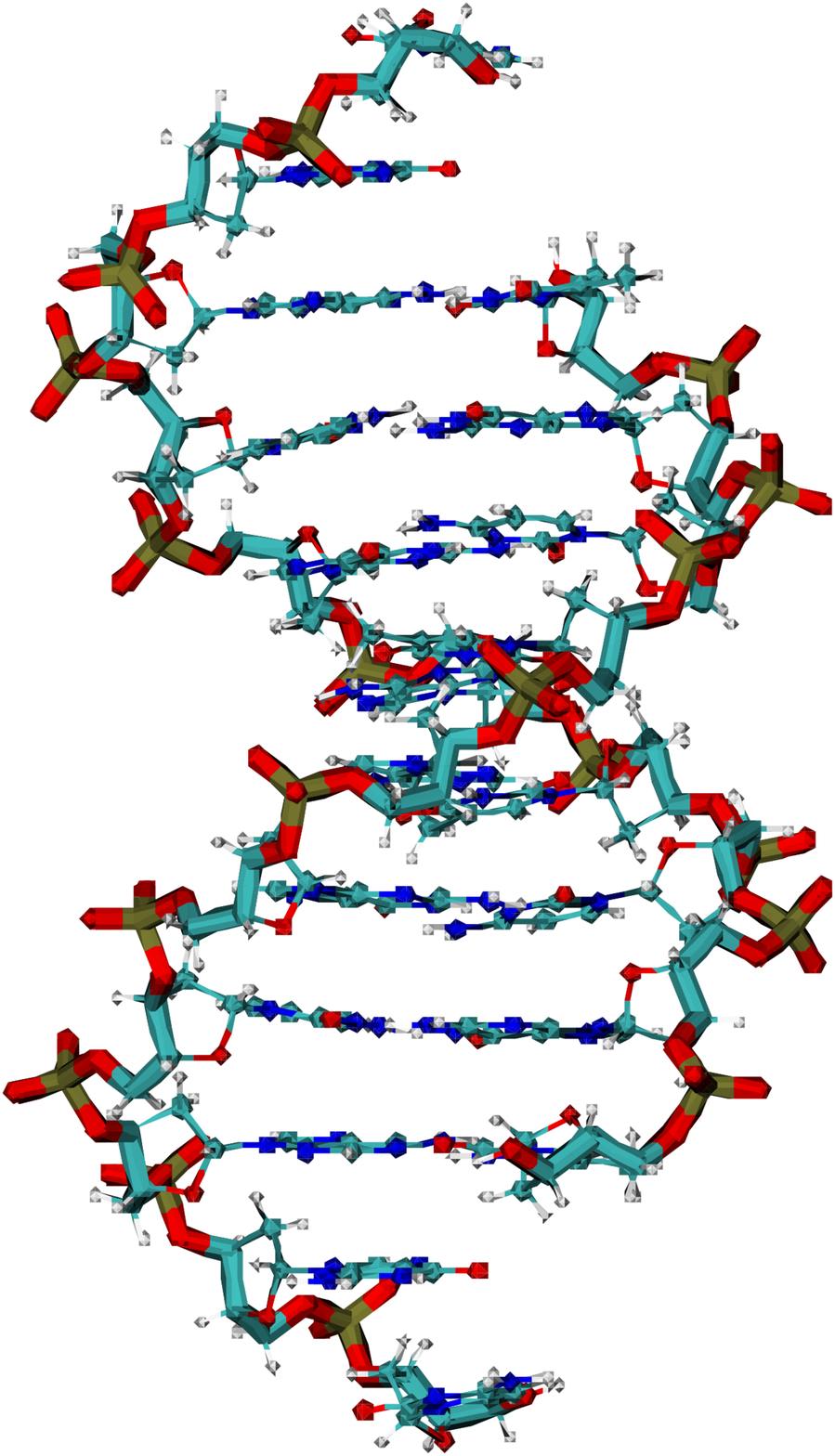}
}
\caption{(Color online) Representation of the B-DNA molecule from the Protein Data Bank \cite{Berman2000} with PDB ID 309D \cite{Qiu1997}. Thick lines represent the backbone bonds whose energy is monitored during the shock wave simulations (see section \ref{sec:results}).}
\label{fig:309D}       
\end{figure}

In the present study we have chosen the B-DNA molecule from the Protein Data Bank \cite{Berman2000} with PDB ID 309D \cite{Qiu1997}. This molecule, represented in Fig. \ref{fig:309D}, consists on a B-DNA duplex containing eight base-pairs and two free nucleotides at each 5'-end. These sticky ends are complementary, forming continuous 10-fold double helix molecules. This molecule is very convenient since it allows the building of DNA duplex models as long as desired, by the replication and displacement of the original molecule as many times as wanted. It opens, as well, the possibility of being used as building block for more complex DNA structures if desired.
\end{sloppypar}

%

\begin{figure*}
\resizebox{0.99\textwidth}{!}{%
  \includegraphics{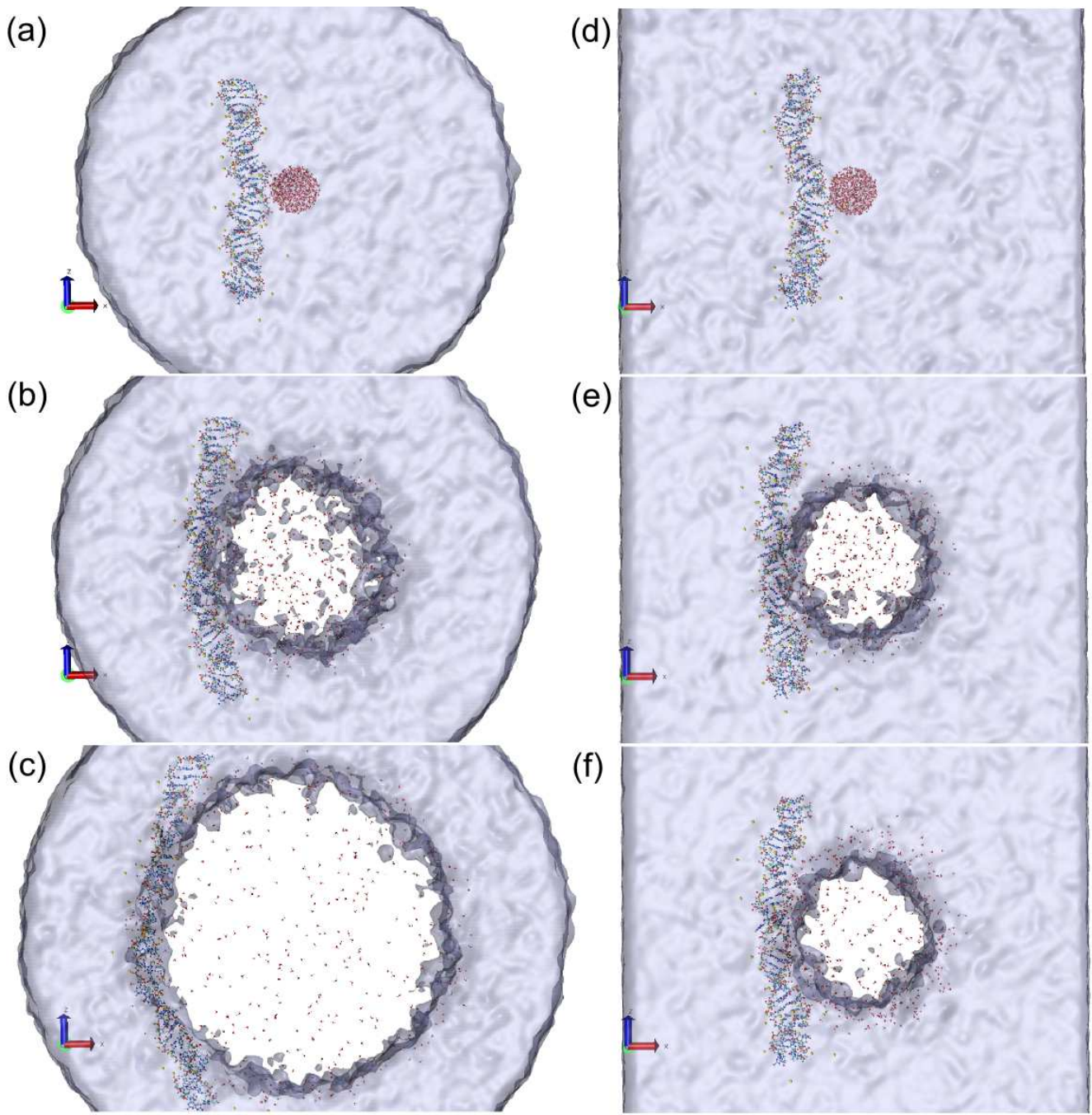}
}
\caption{(Color online) Snapshots of a shock wave induced by a carbon ion in the Bragg peak region in liquid water with the track ($y$ direction) crossing systems I and II at 1 nm distance from the DNA duplex (oriented in the $z$ direction). Panels (a), (b), and (c) correspond to the system I at 0, 5, and 10 ps after the ion traversal, while panels (d), (e), and (f) represent the same times for the system II. The water molecules initially excited by the ion track as well as the DNA duplex are highlighted, with their atoms shown as explicit atoms.}
\label{fig:trajectories}       
\end{figure*}

Currently we have used the 309D molecule to construct a linear DNA duplex, each strand being 30-base long, which we consider is enough for an initial assessment of the shock wave effects in DNA. After building this molecule from the original one from the Protein Data Bank it is solvated in liquid water and an atmosphere of 60 sodium counterions is placed around the DNA (since each base carries one negative charge) by means of the software VMD \cite{Humphrey1996}. Then, using MBN Explorer \cite{Solovyov2012}, the system is optimized by a velocity quenching algorithm, and then equilibrated at $T = 310$ K (body temperature), using the Langevin thermostat with damping time $\tau_{\rm d} = 0.2$ ps, a simulation time step ${\rm d}t = 1$ fs, periodic boundary conditions, the particle mesh Ewald algorithm for the long-range Coulomb interactions, and a van der Waals cut off distance of 13 \AA. 
The geometry of the resulting DNA duplex, as well as the structure of the sodium ions environment, has been checked and compared with reference data \cite{Heinemann1992,Pan2004,Robbins2013}. The equilibrated DNA duplex has then been put in the different water boxes described above for the subsequent shock wave simulations. The DNA duplex, oriented in the $z$ direction, is always placed with its center at 2 nm distance from the track, in a way in which the closest DNA atom is at $\sim 1$ nm from the track. The initial configurations for systems I and II are shown in Figs. \ref{fig:trajectories}(a) and (d).

%




\subsubsection{Setting up the initial conditions of the shock wave}
\label{sec:initial}


An energetic ion losses its energy mainly by electronic excitations and ionizations. As a result a large number of low energy electrons are produced (below 50 eV in the Bragg peak) \cite{Surdutovich2014,deVera2013,deVera2013b,Obolensky2008} which will propagate on the nanometer scale, being stopped very quickly. The dynamics of such a process was studied in Ref. \cite{Surdutovich2015}, where it was shown how the radial dose around the ion track is built up in $\sim 50$ fs. Also, almost all the energy lost by the ion is deposited within 1 nm from the ion track. It is well known that the electron-phonon coupling, the mechanism by which the energy deposited by the secondary electrons can be dissipated, occurs in times much longer than fs, i.e., in the sub-ps scale. This means that a large amount of energy will be locally deposited within $\sim 1$ nm very quickly ($\sim 50$ fs) and that it will be released at once, putting the initial conditions for the formation of the shock wave.

In terms of the molecular dynamics simulations we select the water molecules initially present within 1 nm radius from the ion track. These molecules are highlighted in Fig. \ref{fig:trajectories}, where their atoms are shown as spheres. All the energy lost by the ion (which is not considered explicitly in the simulations since it crosses the system in much shorter times) will be transferred to these molecules
 so the velocities of their atoms (obtained from previous equilibration) are multiplied by a factor $\alpha$, in a way in which their total kinetic energy after the ion crosses the system is:
\begin{equation}
\sum_i^N \frac{1}{2}m_i(\alpha\cdot v_i)^2 = \frac{3Nk_{\rm B}T}{2}+\left| \frac{{\rm d}T}{{\rm d}s} \right| \cdot l \mbox{ . }
\label{eq:Edepos}
\end{equation}
The first term in the right hand side of the equation corresponds to the initial kinetic energy of the excited cylinder (with $N$ atoms) at equilibrium ($T = 310$ K). The second term is the energy lost by the ion when crossing the system, which is its stopping power, $\left|{\rm d}T/{\rm d}s \right|$ times the length of the simulation box, $l$. The simulation of the shock wave is done as indicated previously for the equlibration but without thermostat and with lower values of the time step ${\rm d}t$, depending on the simulation, to ensure conservation of energy.


\section{Results and discussion}
\label{sec:results}

We will start our study of the effects of ion-induced shock waves in DNA duplexes with a carbon ion in the Bragg peak region (with kinetic energy of $\sim$300 keV/u), carbon being a common choice in modern ion beam therapy centers. The stopping power of a carbon ion in the Bragg peak is 900 eV/nm. Figures \ref{fig:trajectories}(a), (b), and (c) show three snapshots of the evolution of the system after irradiation (0, 5, and 10 ps, respectively) when using system I, i.e., a 10 nm radius disc in vacuum (see Table \ref{tab:systems}). The same results are shown in Figs. \ref{fig:trajectories}(d), (e), and (f) for system II, i.e., a parallelepiped of 10 nm track-to-boundary distance with periodic boundary conditions (PBC).

The explosion of the system is much more violent for system I than for system II with the structure of the DNA duplex being heavily distorted in Fig. \ref{fig:trajectories}(c). This is due to the high pressures produced by the shock wave that force the system to expand into vacuum in system I. However, this behavior does not occur in system II, apparently because the periodic boundary conditions provide the pressure to damp the effects of the shock wave. The final structure of the DNA duplex is distorted in system II (Fig. \ref{fig:trajectories}(f)) but not that much as in the case of system I. Therefore, periodic boundary conditions are needed to suppress the system explosion if a small system is used for the simulations.


%
\begin{figure}
\resizebox{1.0\columnwidth}{!}{%
  \includegraphics{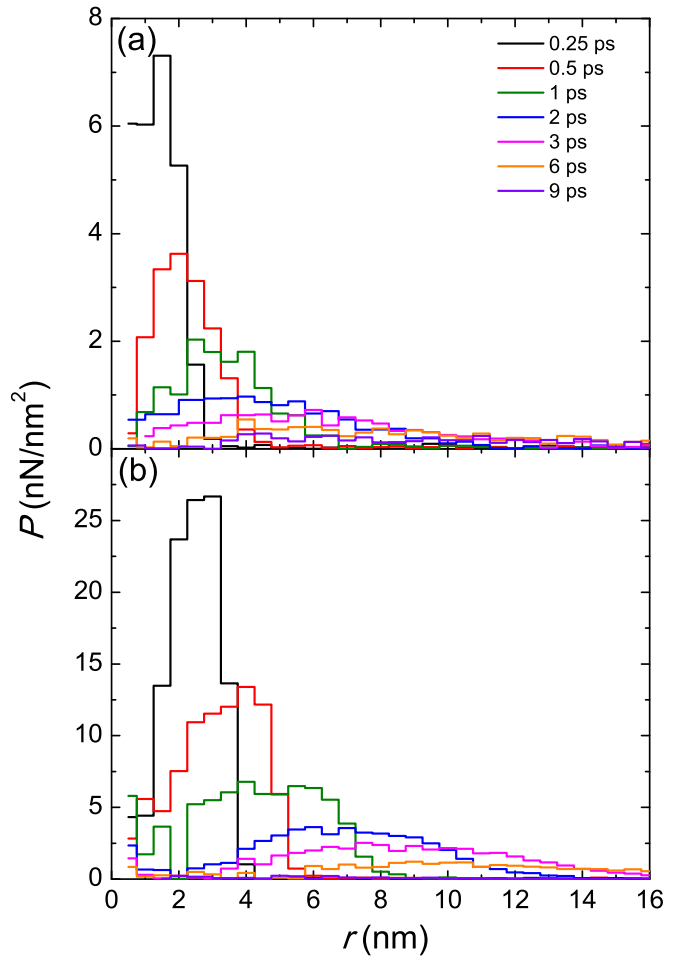}
}
\caption{(Color online) Time evolution of the pressure, as a function of the radius from the track $r$, generated by (a) a carbon ion and (b) an iron ion in the Bragg peak region in system IV, a parallelepiped with 17 nm track-to-boundary distance and periodic boundary conditions (see Table \ref{tab:systems}).}
\label{fig:pressureCFe}       
\end{figure}

A better understanding of this situation can be achieved by calculating the pressure generated by the shock wave. These values can be compared to the results provided by the analytical model (subsection \ref{sec:hydrodynamics}).
To calculate the pressure, virtual walls have been placed at different radial distances $r$ from the track. At several times $t$ during the simulation the number of atoms crossing this wall in each direction has been monitored and their momentum $p_i = m_i v_i$ calculated. The pressure is calculated as:
\begin{equation}
P = \frac{{\rm d}p}{{\rm d}t\,A} = \frac{2 \left(\sum_i p_i-\sum_j p_j\right)}{{\rm d}t\,A} \mbox{ , }
\label{eq:pressure}
\end{equation}
where ${\rm d}t$ is the time passed between frames in the simulation, $A$ is the surface of the cylindrical wall and the indexes $i$ and $j$ refer to the atoms crossing the wall in the outer and in the inner directions respectively. The factor 2 comes from the assumption that, if the wall was there to measure the pressure, the atoms would have an elastic collision coming back after the collision with the same momentum $p_i$ but in nearly opposite direction, so the momentum transfer would be ${\rm d}p_i \simeq 2p_i$.

%
\begin{figure}
\resizebox{1.0\columnwidth}{!}{%
  \includegraphics{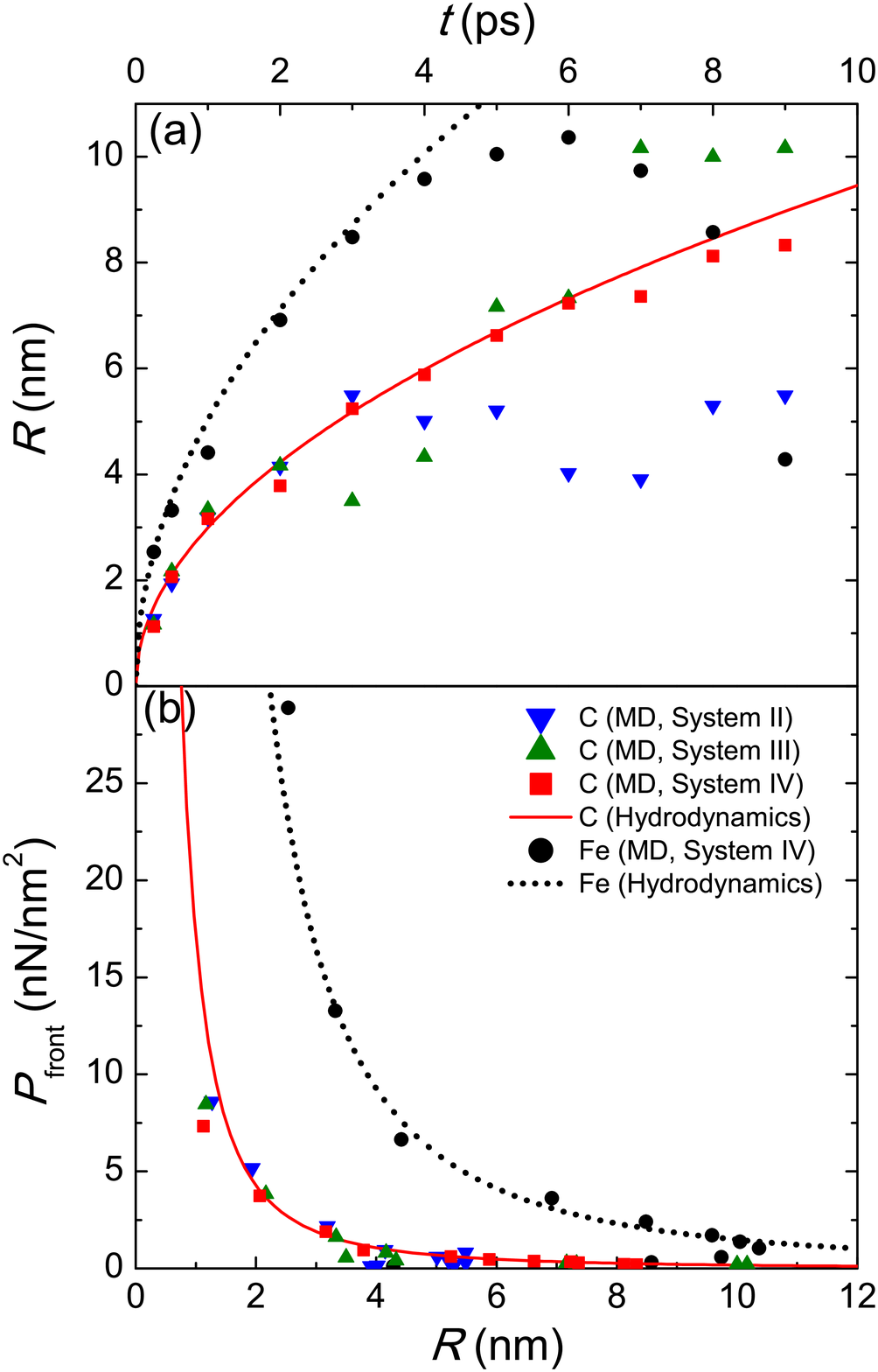}
}
\caption{(Color online) (a) Time evolution of the wave front position for carbon  and iron ion induced shock waves in the Bragg peak region. (b) Pressure of the wave front as a function of its radial position, for carbon and iron induced shock waves in the Bragg peak region. Symbols represent molecular dynamics results, while lines are the predictions of the analytical hydrodynamics model. See the text for further details.}
\label{fig:front}       
\end{figure}

The results for the time evolution of the pressure wave produced by a carbon ion in the Bragg peak in system IV (where boundary effects are not expected) are shown in Fig. \ref{fig:pressureCFe}(a). The wave front can be clearly identified during the first picosecond, where the Gaussian shape is very visible, with a sharp maximum, that propagates rapidly (faster than $\sim$1600 m/s) 
in the radial direction. After the first few picoseconds the pressure wave widens and loses intensity quickly. This time evolution can be compared with the results of the hydrodynamics model (subsection \ref{sec:hydrodynamics}). This comparison is shown for the position of the front and its pressure, respectively, in Figs. \ref{fig:front}(a) and (b), where lines are the results from the analytical model and symbols represent the results from the simulations (which have been obtained by finding the maximum of the Gaussian fitting each curve in Fig. \ref{fig:pressureCFe}(a)). 

The first important observation is that simulation results in system IV follow the results of the analytical model for carbon. However, artifacts appear in the results of systems II and III due to their small size or the lack of PBC. In the case of system II, where the track-to-boundary distance is just 10 nm, the tail of the pressure wave reaches the boundary after $\sim$3 ps. This produces a reflection of the pressure wave in the boundary, which stops the front propagation. Even though the disc-like system III has a track-to-boundary distance of 17 nm, the lack of PBC seems to produce an unnatural propagation of the front at longer times. Neither of these effects are observed for carbon when system IV is used, what suggests that the system has to be large enough and PBC should be applied in order to properly contain the shock wave.



%
\begin{figure}
\resizebox{1.0\columnwidth}{!}{%
  \includegraphics{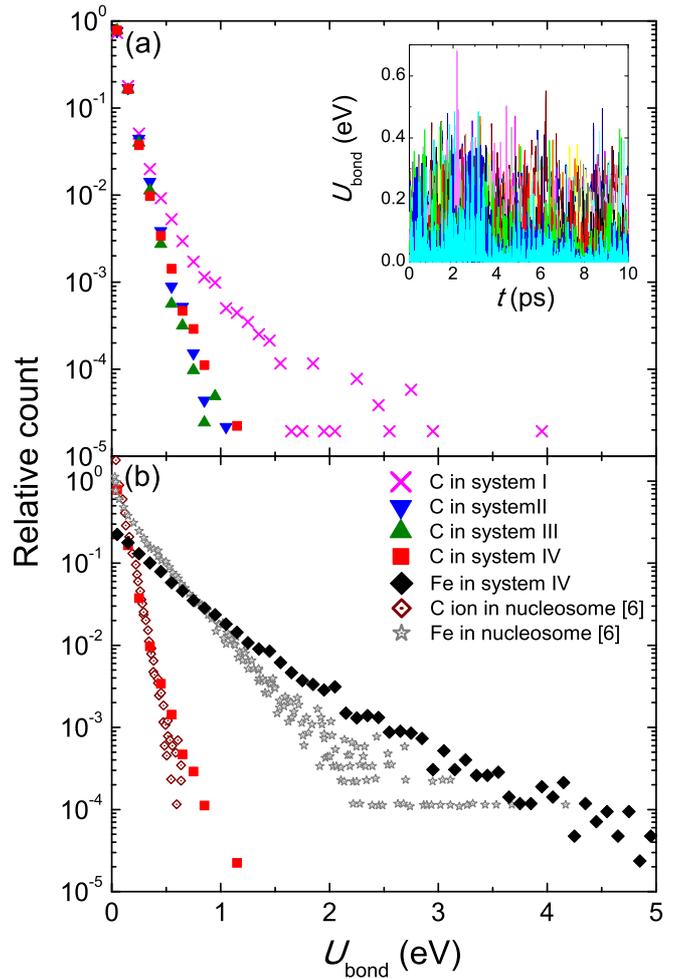}
}
\caption{(Color online) Relative count of energy deposition events in the DNA backbone bonds located within 2.2 nm of the ion track. (a) Present results for a carbon ion in the Bragg peak region in systems I--IV. (b) Present results for carbon and iron ions in system IV, compared with previous results in the nucleosome \cite{Surdutovich2013}. The inset shows the time evolution of the potential energy of some of the bonds analyzed for a carbon ion in system IV.}
\label{fig:Ubond}       
\end{figure}

Before discussing results for a heavier ion (iron), we will analyze the effects of the carbon-induced shock wave in the DNA duplex. We have monitored the energy stored in the covalent bonds of the DNA backbone closest to the track \cite{Surdutovich2013}. The inset in Fig. \ref{fig:Ubond}(a) shows the potential energy of some of the covalent bonds (first term in the right hand side of Eq. (\ref{eq:MM})), initially located within 2.2 nm from the track, as a function of time. The energy stored in the bonds varies in time, with some sudden jumps due to the exposure of the bonds to the pressure of the shock wave. Each of those local maxima can be stored as an energy deposition event. The frequency count of such events for the shock wave produced by a carbon ion in the Bragg peak in all the systems I-IV is represented in Fig. \ref{fig:Ubond}(a). This frequency, which presents an exponential behavior, can be parameterized to estimate the probability for inducing single strand breaks, regarded as energy deposition larger than some given threshold, typically around 2.5 eV \cite{Surdutovich2013}. Clearly, system I presents a very different behavior while systems II-IV converge. This result arises from the artificial violent explosion of system I, due to the lack of pressure to damp the shock wave. The convergence between systems II-IV emphasizes the importance of the first picoseconds of the shock wave in terms of DNA damage: even though the pressure wave differs in these systems after $\sim 3$ ps, their energy deposition profile is similar, highlighting the fact that it is built up during the first picoseconds after the explosion. The energy deposition profile is compared with previous simulations in the nucleosome \cite{Surdutovich2013} in Fig. \ref{fig:Ubond}(b), where we find a fairly good match, allowing for the different geometries.

\begin{figure}
\resizebox{1.0\columnwidth}{!}{%
  \includegraphics{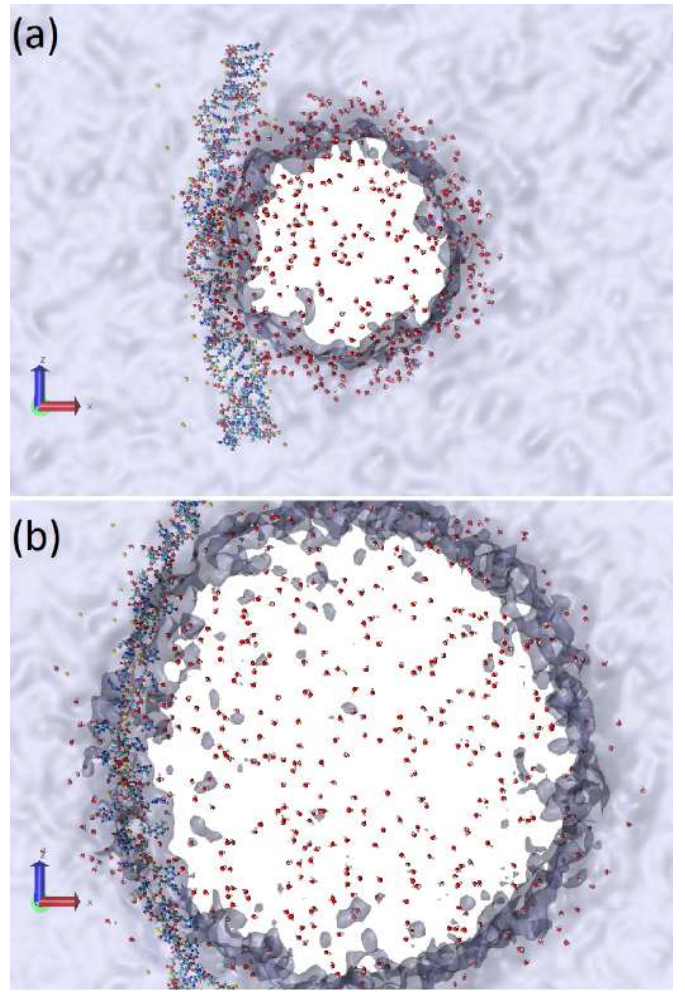}
}
\caption{(Color online) Geometry of system IV, 10 ps after passage of (a) a carbon ion and (b) an iron ion, in the Bragg peak region. Nanochannels are formed in liquid water, and the DNA structure is distorted, particularly heavily in the case of the iron ion.}
\label{fig:C_Fe_final}       
\end{figure}

From these simulations it is clear that there is a relatively low probability of producing single strand breaks by the shock wave in the case of carbon ions (assuming a SSB threshold of 2.5 eV; however, such thresholds can be even lower \cite{Surdutovich2013}). In terms of a possible experimental verification of these results it would be convenient to study ions with larger stopping powers since these can produce larger numbers of strand breaks, so they are easier to detect. For this reason we have performed a simulation of the shock wave produced by an iron ion in the Bragg peak region, having a stopping power of 7195 eV/nm. System IV has been used for this simulation simulation since, from the results for carbon, the other systems would be too small. The evolution of the pressure wave for iron is shown in Fig. \ref{fig:pressureCFe}(b), the evolution of the front is depicted in Fig. \ref{fig:front}, and the energy deposition profile in the DNA backbone bonds is shown in Fig. \ref{fig:Ubond}(b). The final geometry of the system after 10 ps, as compared with the carbon case, is shown in Fig. \ref{fig:C_Fe_final}.


Both the position of the front and its pressure, shown in Fig. \ref{fig:front}, seem to follow the predictions of the analytical model but only up to 4 ps. After that, it can be clearly seen that the front actually goes backwards. This behaviour is caused by the reflection of the pressure wave at the boundaries of the system, as it happened for carbon with system II. As expected from Eqs. (\ref{eq:R_t}) and (\ref{eq:P_R}), the velocity of propagation of the iron shock wave is 1.682 faster than for carbon and the pressure of the front is 8 times larger. This results from the fact that the stopping power for iron in the Bragg peak region is 8 times larger than for carbon. 

Such large pressures lead to larger distortions of the DNA duplex being observed for iron in comparison with carbon in Fig. \ref{fig:C_Fe_final}. This leads to a larger number of high energy deposition events in DNA backbone bonds, as shown in Fig. \ref{fig:Ubond}(b), where the slope of the curve is much smaller than for carbon ion and many more events larger than 2.5 eV (a conservative estimation for the threshold for single strand break production) are produced. This would justify the use of heavier ions for a possible experimental verification of shock wave effects, where more single strand breaks would be detected. The results for iron are compared with previous simulations performed for nucleosome \cite{Surdutovich2013}. In this case, the present results are not that close to the previously reported values. However, the order of magnitude seems to be similar. Such differences could be due to several reasons, the most probable being the different geometry. The larger slope of the previously reported data suggests that the histone protein has some protective role in DNA damage. In the present work the DNA duplex is free, allowing a larger stretching of the molecule which might be impeded in the nucleosome. Also the fact that system IV is not large enough for the iron-induced shock wave could have some contribution. However, as seen from the results for carbon, the first picoseconds of the simulation are the most relevant for the DNA damage and the molecular dynamics simulation follows properly the analytical results up to 4 ps. It is clear, in any case, that a larger system is needed for the simulation of an iron ion shock wave. This has not been done here in order not to increase the computational expense of the calculations. However, precautions should be taken in future calculations to optimize the  box for each ion. In any case, the main conclusion that can be extracted from Fig. \ref{fig:front} is that molecular dynamics simulations can reproduce perfectly the results predicted by the analytical hydrodynamics model as long as the system used for the simulations is large enough for containing the wave and PBC are applied. Indeed the hydrodynamics model can be used as a consistency check to analyze whether the system built for molecular dynamics simulation is appropriate or not.

%

\section{Conclusions}
\label{sec:conclusions}

In this paper we have presented a molecular dynamics study of the effects of ion-induced shock waves in biological media. The study focuses on the pressure waves arising from the heating of liquid water after energetic carbon and iron ions traversal (in the Bragg peak region) and on the effects of such waves on short DNA duplexes. 


\begin{sloppypar}
The first important conclusion of this paper is that molecular dynamics simulations of ion-induced shock waves can reproduce the results obtained by the analytical hydrodynamics model previously reported, in terms of front velocity and pressure, as long as the simulation box is properly designed. This point is of great importance: ion-induced shock waves have not been detected yet experimentally and their existence has been so far only predicted theoretically. The fact that independent techniques, such as classical molecular dynamics and classical hydrodynamics, coincide in the prediction of the properties of the shock waves adds arguments in favor of their existence and establishes the theoretical grounds over which possible experiments can be designed and interpreted for their confirmation.
\end{sloppypar}

Molecular dynamics simulations reproduce the analytical results only if the system is well designed. This means that it has to be large enough to contain the pressure waves produced and that preferably periodic boundary conditions should be applied to avoid boundary effects in the simulation, such as the reflection of the pressure wave in the periodic boundaries or the expansion of the system into vacuum when PBC are not applied. In the present case, a track-to-boundary distance of 17 nm has been demonstrated to be large enough for carbon-induced shock waves in the Bragg peak region. For iron larger boxes should be used in future work. However, smaller boxes can be used for certain simulations as long as one is interested in short times and distances from the track where most of the damage occurs. The analytical hydrodynamics model can be always used to check the consistency of the results, as shown in this paper. This is something that has not been considered before and has not been checked in other works.


The energy deposition profile in the DNA backbone bonds has been analyzed, both for carbon and iron ions, as an estimation of possible biological damage. The results obtained for carbon are very similar to those reported previously for molecular dynamics simulations in the nucleosome \cite{Surdutovich2013}. For the case of iron ions we have found some differences with the previous published data. However, the order of magnitude of the results is quite similar, confirming the systematics of the stopping power of the ion on the expected damage of DNA. The observed differences are most likely due to the different geometry of the systems used, where the presence of the histone protein in the nucleosome can have some protective effect.

The present results establish a solid procedure for performing more systematic simulations of shock wave effects in DNA duplexes, where the analytical hydrodynamics model can be used as a benchmark. The properties of the shock waves for ions of different stopping powers can be predicted and their effects on DNA determined. This opens the door to new simulation improvements, such as the inclusion of reactive force fields for a better prediction of DNA damage \cite{Bottlander2015,Sushko2016} or the study of the shock wave effects in the propagation of secondary species generated around ion tracks. Such systematic studies will allow a better understanding of the biological relevance of ion-induced shock waves and will be useful for the future design and interpretation of potential experiments for their detection.



\begin{acknowledgement}
\begin{sloppypar}
The authors acknowledge financial support from the European Union’s FP7-People Program (Marie Curie Actions) within the Initial Training Network No. 608163 "ARGENT", Advanced Radiotherapy, Generated by Exploiting Nanoprocesses and Technologies. The molecular dynamics simulations were performed at the computer clusters IMACT, at The Open University (UK), and KELVIN, at Queen's University Belfast (UK).
\end{sloppypar}
\end{acknowledgement}

%
%

\end{document}